\documentstyle[12pt,epsf]{article}
\newcommand{\beq}{\begin{equation}}
\newcommand{\eeq}{\end{equation}}
\newcommand{\beqa}{\begin{eqnarray}}
\newcommand{\eeqa}{\end{eqnarray}}
\newcommand{\boldtau}{\mbox{\boldmath $\tau$}}

\newcommand{\boldpi}{\mbox{\boldmath $\pi$}}

\begin{document}

\begin{titlepage}

\hfill{KRL MAP-265}

\vspace{2.5cm}

\begin{center}
{\Large\bf The Anapole Form Factor of the Nucleon}

\vspace{2.0cm}

{\bf C.M. Maekawa}\footnote{{\tt maekawa@krl.caltech.edu}}
and 
{\bf U. van Kolck}\footnote{{\tt vankolck@krl.caltech.edu}}

\vspace{0.8cm}
{\it
 Kellogg Radiation Laboratory, 106-38 \\
 California Institute of Technology \\
 Pasadena, CA 91125}
\end{center}

\vspace{1.5cm}

\begin{abstract}
The anapole form factor of the nucleon is calculated in
chiral perturbation theory in leading order.
To this order, the form factor originates from
the pion cloud, and is proportional to
the non-derivative parity-violating pion-nucleon coupling.
The momentum dependence of the form factor 
---and in particular, its radius--- is completely
determined by the pion mass. 
\end{abstract}

\vspace{2cm}
\vfill
\end{titlepage}

\setcounter{page}{1}

There has been much recent interest in parity-violating electron
scattering on the nucleon and light nuclei
as a source of information about nucleon structure,
in particular strangeness content.
The SAMPLE collaboration has measured the 
longitudinal electron asymmetry for scattering on the
proton \cite{sample} and the deuteron
at $Q^2=-q^2=0.1$ GeV$^2$, where $q$ is the
momentum transferred to the target.
Experiments at JLab (HAPPEX \cite{happex} and G0 \cite{g0})
extend these measurements to higher
$Q^2$.

A class of contributions to the measured electron asymmetry
comes from the anapole form factor of the nucleon.
The anapole is a parity-violating electromagnetic moment 
of a charged particle with spin, 
related at classical level to  
the magnetic field induced within a torus
by a winding wire \cite{zeldovich}.
It vanishes for on-shell photons and thus cannot
be separated from contact electron-particle operators.
Nevertheless, in the case of hadrons at low energies,
which can be formulated as an effective field theory (EFT),
the anapole does constitute a class of contributions
that is independent of the choice of photon (gauge),
nucleon and pion fields, and therefore can be examined in itself.
The anapole contribution to the asymmetry needs to be
understood before one can draw definitive conclusions
about the strangeness current.

Here we focus on the anapole form factor of the nucleon
at $Q\ll M_{QCD}$, where $M_{QCD}\sim 1$ GeV is the typical
mass scale in QCD. In this kinematic regime
we can perform a systematic expansion of the form factor
in powers of $Q/M_{QCD}$
(times functions of $Q/m_\pi$).
This expansion receives 
analytic contributions from short-range physics (rho mesons, etc.) 
and non-analytic contributions from the long-range pion cloud.
The latter are calculable in a model-independent way 
in chiral perturbation theory ($\chi$PT).
We discuss the relative order
these contributions appear, and show that
to lowest order {\it only} the pion cloud contributes. 
The anapole form factor at this order is then 
a calculable function of $(Q/m_\pi)^2$,
completely determined by pion properties and
pion-nucleon couplings in principle known from other processes.

The anapole form factor at $Q=0$ (the anapole moment)
has been calculated before, together with subsets
of sub-leading contributions \cite{holmus,kapsav}, 
and depends on the leading, yet poorly-determined, parity-violating
pion-nucleon coupling.
However, it is not the anapole moment {\it per se}
that is relevant for the above scattering experiments. 
Because it is generated by the pion cloud, 
the scale that governs the momentum dependence of the form factor is
set by $m_\pi$, and not $M_{QCD}$. In the chiral limit, the form factor
would vary wildly between  $Q=0$ and $Q=M_{QCD}$,
but even in the real world it could
be quite different (say, by a factor of a few) at $Q\sim 300$ MeV 
(SAMPLE experiment) than at $Q=0$.
For the proper interpretation of the new experimental data one thus
needs to investigate the momentum dependence of the anapole contribution.

We present here 
the leading-order results for the nucleon anapole form factor,
which turns out to be purely isoscalar.
We recover the known result for the anapole moment \cite{holmus,kapsav}, 
and further predict its momentum dependence,
for which we give an analytic expression.
This momentum dependence is given by the known pion mass
and is an example of a low-energy theorem.
In particular, the radius of the anapole form factor is,
as expected, $\propto 1/m_\pi$ and blows up in the chiral limit. 
These predictions could in the future be tested.
We will also show that
dimensionless factors are such that the scale of the $Q^2$ variation
is $\sim 6m_\pi$, numerically (but not parametrically) close to
the rho mass, $m_\rho$. As a consequence,
the form factor at this order does not display much variation
with momentum.

Here we denote by $iJ^\mu_{an}$ the 
parity-violating nucleon current
that interacts
with the electron current $-ie \, \overline{e} \gamma^\mu e$ via the 
photon propagator $iD_{\mu\nu}=- i(\eta_{\mu\nu}/q^2 +\ldots)$ 
to produce a contribution
\begin{equation}
iT= -i e \, \overline{e}(k') \gamma^\mu e(k) \, D_{\mu\nu}(q) 
        \, \overline{N}(p') J^\nu_{an}(q) N(p),
\end{equation}
to the electron-nucleon $S$ matrix.
We have $q^2=(p-p')^2\equiv -Q^2<0$.

We want to evaluate $J^\mu_{an}$ at $Q\ll M_{QCD}$.
Because the nucleon mass is heavy,
$m_N\sim M_{QCD}$, in this regime the nucleon is essentially
non-relativistic, and is parametrized by a given
velocity $v^\mu$ and spin $S^\mu$
($S^\mu=(0, \vec{\sigma}/2$) in the nucleon rest frame
where $v^\mu=(1, \vec{0}$)).
As we are going to see, we can write
\begin{equation}
J^\mu_{an}(q)= \frac{2}{m_N^2} 
               \left(a_0 F^{(0)}_A(-q^2)
                +a_1 F^{(1)}_A(-q^2) \tau_3\right) (S^\mu q^2- S\cdot q q^\mu),
\label{cur}
\end{equation}
where $F^{(i)}_A(0)=1$ and $\tau_i$ is the $i$th Pauli matrix in isospin
space. 
$a_0$ ($a_1$) is the isoscalar (isovector) anapole moment 
of the nucleon
and $a_0 F^{(0)}_A(Q^2)$ ($a_1 F^{(1)}_A(Q^2)$)
the corresponding form factor.

At $Q\sim m_\pi$, the resolution of the virtual photon
is enough to see pions, and
pionic contributions have to be 
taken into account explicitly.
Because the delta-nucleon mass difference is comparable
to the pion mass, the delta isobar is also
a relevant low-energy degree of freedom.
All these contributions can be calculated in a model-independent way starting
from the most general Lagrangian involving nucleons $N$, pions $\boldpi$
and deltas that transforms under the symmetries of QCD
in the same way as QCD itself.
This EFT has been described countless times
in the literature; here we 
only highlight the features strictly
relevant to our calculation, and refer the reader
to a good review ---such as Ref. \cite{ulfreview}--- for details.

Since the symmetries allow an infinite number of interactions,
it is imperative to have an ordering scheme for the various contributions.
Chiral symmetry plays a fundamental role here.
Such a power counting argument is possible because all
strong interactions bring in a small scale.
Interactions that preserve chiral symmetry involve derivatives of the pion
field, so they bring to amplitudes powers of the small momentum,
or powers of the delta-nucleon mass difference;
and interactions
that break chiral symmetry involve the quark masses,
so they bring powers of the pion mass.
One can then order the strong interactions in the Lagrangian,
${\cal L}=\sum_\Delta {\cal L}^{(\Delta)}$,
according to the chiral index $\Delta \equiv d+n/2 -2$,
where $d$ is the sum of the number of derivatives and
powers of the pion mass and of the delta-nucleon mass difference,
and $n$ is the number of fermion fields \cite{weinberg}. 
Electromagnetic interactions also break chiral symmetry;
they do not necessarily contain quark masses,
but they are proportional to the small charge $e$.
It is convenient to account for factors of $e$
by enlarging the definition of $d$ accordingly.
It can then be shown that in leading order in the strong and electromagnetic
interactions, contributions come from
\begin{equation}
 {\cal L}^{(0)}_{str/em}  = 
          \frac{1}{2}(D_\mu\boldpi )^{2}
          -\frac{1}{2}m_{\pi}^{2}\boldpi^{2} 
          +\overline{N}iv\cdot D N 
          -\frac{g_{A}}{f_{\pi}} 
          \overline{N}(\boldtau\cdot S\cdot D \boldpi)N
          +\ldots
\label{la0}
\end{equation} 
\noindent
Here $f_{\pi}=93$ MeV is the pion decay constant,
$D_\mu= (\partial_\mu -ie Q A_\mu)$
with $Q^{(\pi)}_{ab}=-i \varepsilon _{3ab}$
and $Q^{(N)}=(1+\tau_3)/2$,
and ``\ldots'' stand for interactions with more
pion, nucleon and/or delta fields, that are not needed explicitly
in the following.
Note that at this order the nucleon is static and couples
only to longitudinal photons. Kinetic corrections and magnetic couplings 
have relative size $O(Q/M_{QCD})$ and appear in ${\cal L}^{(1)}_{str/em}$.
The same is true for the delta isobar, including
the nucleon-delta transition through coupling to a
transverse photon.
The parity-conserving pion-nucleon coupling
is given by the axial-vector coupling of the nucleon;
according to naive dimensional analysis $g_{A}=O(1)$, 
which is indeed observed, $g_{A}=1.26$ (see, e.g., Ref. \cite{ulfreview}). 
A term in ${\cal L}^{(2)}_{str/em}$ provides an 
$O((m_\pi/M_{QCD})^2)$ correction
that removes the so-called Goldberger-Treiman discrepancy.

Weak interactions in this EFT have been discussed in
Ref. \cite{kapsav}. 
The four-fermion interactions between quarks generated
by $W$ and $Z$ exchange break chiral
symmetry with a strength given by the Fermi constant 
$G_F=1.17 \cdot 10^{-5}$ GeV$^{-2}$.
The effective operators it entails
at low energies are proportional to the Fermi constant times
the square of a mass scale.
A natural scale is the pion decay constant,
so we assume that these operators have coefficients of order of
$G_F f_\pi^{2}\sim 10^{-7}$ times other natural scales
\footnote{There are clearly factors of $4\pi$ that might
not be accounted for properly this way, but they will
not affect the {\it relative} order of weak interactions.}.
There is one
pion-nucleon interaction with negative index,
\begin{equation}
{\cal L}^{(-1)}_{weak}=
-\frac{h_{\pi NN}}{\sqrt{2}} \overline{N} (\boldtau\times\boldpi)_3 N
+\ldots
\label{law-1}
\end{equation}
Here $h_{\pi NN}$ is the parity-violating pion-nucleon coupling.
Again, according to naive dimensional analysis
we expect $h_{\pi NN} f_\pi = G_F f_\pi^{2} M_{QCD}$,
or $h_{\pi NN}=O(G_F f_\pi M_{QCD}) \sim 10^{-6}$.
The value of $h_{\pi NN}$ is not well determined
(for a review of constraints 
on parity-violating parameters, see, for example, 
Ref. \cite{hpiNN}
\footnote{At lowest order, 
our parameters relate to those of Ref. \cite{DDH} (DDH) by
$h_{\pi NN}g_A/f_\pi= f_\pi^{DDH} g_{\pi NN}^{DDH}/m_N$.}).

There are various weak interactions with indices 0, 1, and 2.
The first pion-nucleon-delta coupling, for example,
has to contain in the fermion rest frame
the $4\times 2$ transition matrix $\vec{S}_{tr}$;
in other words, we cannot
combine nucleon and delta in a spin zero object.
Rotational invariance then demands
the presence of a gradient to form a scalar,
$\vec{S}_{tr}\cdot \vec{\nabla}$.
Such interaction can appear only in ${\cal L}^{(0)}_{weak}$,
and its effects are suppressed by $O(Q/M_{QCD})$ relative
to those stemming from the Lagrangian (\ref{law-1}).
The same reasoning applies to other operators.
Particularly relevant among higher-order interactions is 
\begin{equation}
{\cal L}^{(2)}_{weak}=\frac{2}{m_N^2} 
\overline{N} (\tilde{a}_0+\tilde{a}_1 \tau_3) S_\mu N 
     \, \partial_\nu F^{\mu\nu}
                      +\ldots,
\label{law2}
\end{equation}
because $\tilde{a}_0$ ($\tilde{a}_1$) is a short-range contribution to the
isoscalar (isovector) anapole moment. 
{}From dimensional analysis,
$\tilde{a}_i/m_N^2 =O(e G_F f_\pi^{2}/m_N^2)=O(e G_F/(4\pi)^2)$.
Direct  short-range contributions to the 
momentum dependence of the anapole form factor first appear
in ${\cal L}^{(4)}_{weak}$,
being further suppressed by $O((Q/M_{QCD})^2)$.

\begin{figure}[t]
\begin{center}
\epsfxsize=12cm
\centerline{\epsffile{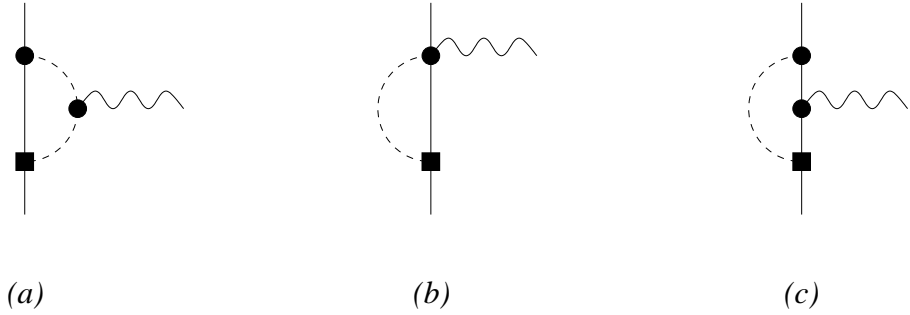}}
\end{center}
\caption{Diagrams contributing to the nucleon anapole form factor in
leading order. Solid, dashed and wavy lines 
represent nucleon, pion, and (virtual) photon,
respectively; circles and squares stand for interactions
from ${\cal L}^{(0)}_{str/em}$ and ${\cal L}^{(-1)}_{weak}$, respectively.
For simplicity only one of two possible orderings are shown here.}
\label{anafig1}
\end{figure}

Let us now consider the sizes of specific contributions
to $J^\mu_{an}(q)$.
The first tree level contribution is given by the vertex
generated by the weak Lagrangian (\ref{law2}).
It has a size $O(e G_F Q^2/(4\pi)^2)$,
and thus contributes $O(e G_F m_N^2/(4\pi)^2)$
to the anapole form factor (specifically, the anapole moment).
The lowest-order one-loop graphs are built out of
one vertex from the weak Lagrangian  (\ref{law-1}) and
all other vertices from the strong Lagrangian (\ref{la0}).
They are depicted in Fig. \ref{anafig1}.
Consider for example the graph in Fig. \ref{anafig1}(a).
The coupling of the photon to the pion contributes
$O(eQ)$, the parity-violating pion-nucleon coupling
$O(G_F f_\pi M_{QCD})$, the parity-conserving 
pion-nucleon coupling $O(Q/f_\pi)$,
each pion propagator $O(1/Q^2)$,
the nucleon propagator $O(1/Q)$, and the loop integration
$O(Q^4/(4\pi)^2)$. 
We expect the whole graph then to be 
$O(e G_F M_{QCD} Q/(4\pi)^2)$,
and contribute  $O(e G_F M_{QCD}m_N^2/((4\pi)^2 Q)$ 
to the anapole form factor. 
One can easily verify that Figs. \ref{anafig1}(b),(c)
give contributions of the same size.
These long-range contributions are $O(M_{QCD}/Q)$ larger than the
most important short-range contribution.
The factor $1/Q$ represents an infrared enhancement.
The anapole moment, which is a constant, is
$O(e G_Ff_\pi^2 M_{QCD}/m_\pi) \sim 5 \times 10^{-7} \, e$, 
and diverges in the chiral limit.
The anapole form factor is $O(e G_Ff_\pi^2 M_{QCD}/m_\pi)$
times a function $F(Q^2/m_\pi^2)=O(1)$.

One can now show that all other contributions 
(including all delta effects) are of higher order.
First, other purely short-range contributions 
start at ${\cal L}^{(4)}_{weak}$, and contribute
at most $O(e G_F Q^2/(4\pi M_{QCD})^2)$ to the form factor.
Second, other long-range contributions 
contribute
at most $O(e G_F Q/((4\pi)^2M_{QCD}))$.
Consider other one-loop graphs: they all will have
at least {\it (i)} one vertex from ${\cal L}^{(1)}_{str/em}$; or 
{\it (ii)} one vertex from
${\cal L}^{(0)}_{weak}$. 
Examples are, respectively, graphs where
{\it (i)} the photon couples to a virtual nucleon
or delta magnetically, or causes a nucleon-delta transition;
and  {\it (ii)} the parity-violating pion coupling is a 
nucleon-delta transition.
Diagrams containing one insertion of either of these couplings
are formally suppressed
by $Q/M_{QCD}$ compared to the leading-order contributions
obtained from the charge coupling present in 
the Lagrangian (\ref{la0}) and the parity-violating
pion-nucleon vertex in (\ref{law-1}): 
they first contribute in sub-leading order.
A consequence is that, although the delta is treated
on the same footing as the nucleon, its 
(virtual) contributions to the nucleon anapole are all
subleading in power counting.
Finally, even the most important
two-loop graphs, made out of  ${\cal L}^{(0)}_{str/em}$
and ${\cal L}^{(-1)}_{weak}$, 
should be suppressed by the usual $(Q/4\pi f_\pi)^2$
associated with loops in $\chi$PT. 

The leading contribution to the anapole form factor comes thus from
the graphs in Fig. \ref{anafig1}.
They are evaluated with
$v \cdot q =0$, as a consequence of the static nature of the nucleon
at this order.
In particular, Fig. \ref{anafig1}(c) vanishes.
It is straightforward to calculate the other diagrams as well.

The other two sets of graphs combine to give a result in the form of 
Eq. (\ref{cur}).
We find a purely isoscalar result,
\begin{eqnarray}
a_0 &=& \frac{eg_A h_{\pi NN}}{48\sqrt{2}\pi} 
        \frac{m_N^2}{m_\pi f_\pi}\label{a0} \\
a_1 &=& 0. \label{a1}
\end{eqnarray}
This result for the anapole moment agrees
with Ref. \cite{kapsav} 
\footnote{The Lagrangian of Ref. \cite{kapsav} can be obtained from ours
by a redefinition of the sign of the pion field.}. 
Ref. \cite{holmus} contains a partial set of
sub-leading contributions
because it pre-dates the heavy-fermion expansion
employed here and in Ref. \cite{kapsav}.
Note that Eq. (\ref{a0})
is $\sim \sqrt{2}\pi/3$ larger than our naive estimate.
Eqs. (\ref{a0}, \ref{a1}) are predictions of $\chi$PT,
but unfortunately the value of $h_{\pi NN}$ is currently
not well determined.

\begin{figure}[t]
\begin{center}
\epsfxsize=12cm
\centerline
{\epsffile{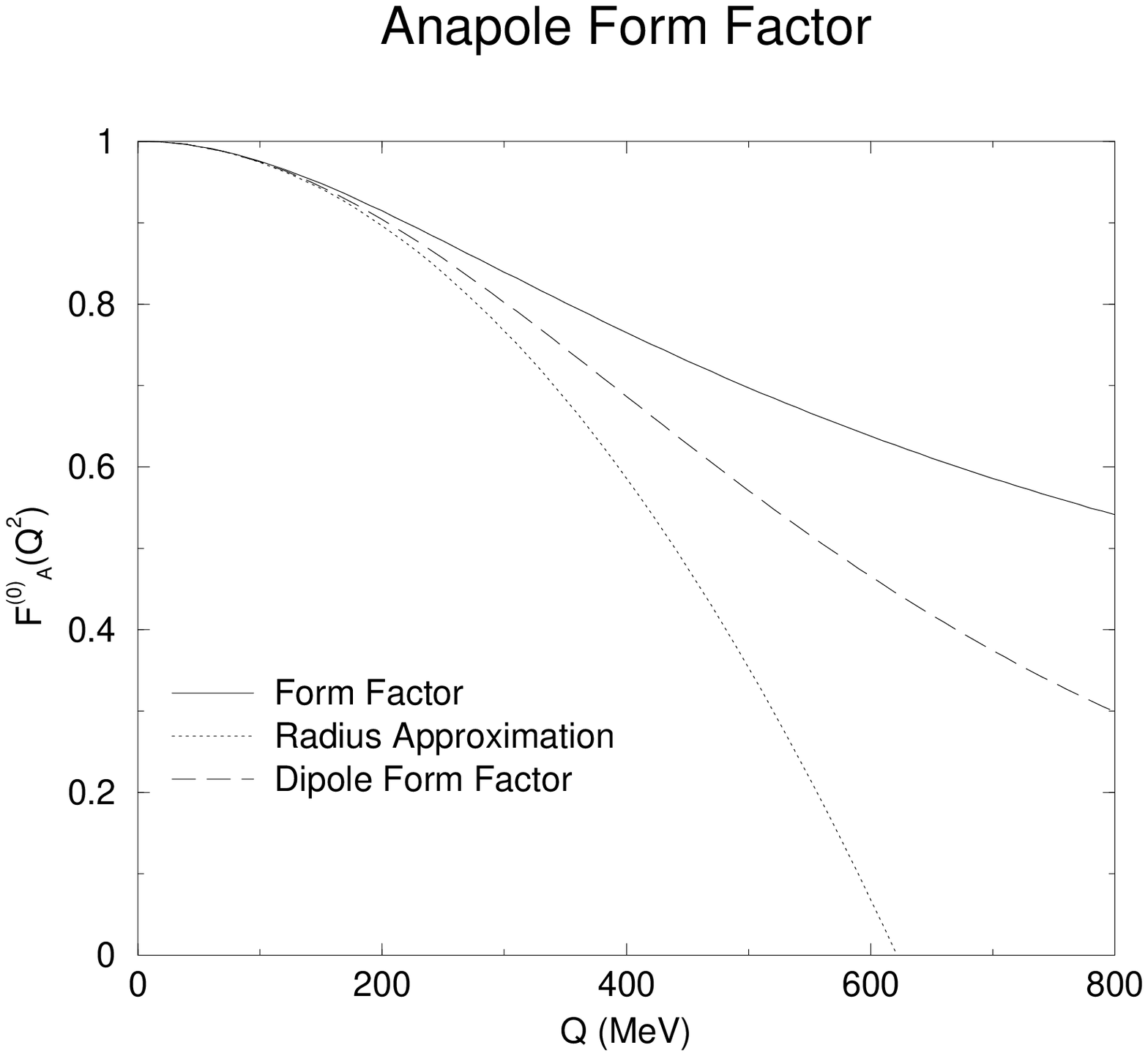}}
\end{center}
\caption{The isoscalar anapole form factor $F^{(0)}_A$
as function of $Q$:
$\chi$PT in leading order, Eq. (\ref{FF}) (solid line);
quadratic approximation, Eq. (\ref{radapp}) (dotted line);
and dipole approximation, Eq. (\ref{dipapp}) (dashed line).}
\label{anafig2}
\end{figure}

Because the form factor is given by lowest-order loop graphs,
it depends on the combination $Q^2/m_\pi^2$ only.
Because all $Q$ dependence comes from Fig. \ref{anafig1}(a),
we find that it actually is a function of $Q^2/(2m_\pi)^2$:
\begin{equation}
F^{(0)}_A(Q^2) = \frac{3}{2}
                 \left\{- \left(\frac{2m_\pi}{\sqrt{Q^{2}}}\right)^{2}
                +\left( 
\left(\frac{2m_\pi}{\sqrt{Q^{2}}}\right)^{2}+1\right) 
\frac{2m_\pi }{\sqrt{Q^{2}}}
\arctan \frac{\sqrt{Q^{2}}}{2m_\pi } \right\}
\label{FF}
\end{equation}
This is also a testable prediction of $\chi$PT:
it is plotted as function of $Q$ in Fig. \ref{anafig2}.

The variation of the form factor with $Q$ is somewhat
smaller than expected. 
This can be seen by looking at the anapole square radius
(defined in analogy to the charge square radius):
\begin{equation}
\langle r_{an}^2 \rangle
= -6 \left. \left(\frac{dF^{(0)}_A}{dQ^2}\right) \right|_{Q^2=0}
= \frac{3}{10m_\pi ^{2}}.
\end{equation}
We see the square radius is smaller than expected by
a factor 5. 
At $Q\ll m_\pi$ we can write
\begin{equation}
F^{(0)}_A(Q^2) =1-\frac{1}{5}\frac{Q^2}{(2m_\pi) ^{2}}
              +O\left( \frac{Q^4}{(2m_\pi)^4}\right).
\label{radapp}
\end{equation}
This radius approximation to the form factor is 
also displayed in Fig. \ref{anafig2}.

The accidentally small square radius 
makes the mass scale that governs the $Q$ dependence
near $Q=0$ larger than $2m_\pi$.
Data on the proton Sachs form factors are well fitted
by a dipole profile with a mass scale
close to $m_\rho$.
A dipole form factor with the correct anapole square radius
is 
\begin{equation}
F^{(0)}_A(Q^2) =\frac{1}{(1+(Q/M)^2)^2}+O\left( \frac{Q^4}{(2m_\pi)^4}\right),
\label{dipapp}
\end{equation}
with the mass scale $M= 2\sqrt{10} m_\pi =880$ MeV,
instead of $2m_\pi$.
$M$ is numerically close to $m_\rho$,
but this is clearly a coincidence that
would be destroyed if the quark masses where much
smaller than what they are.
In fact, short-range contributions
such as that from the rho appear only in the
counterterms at higher orders,
so $M$ has no obvious connection to $m_\rho$.
The dipole approximation is also shown in 
Fig. \ref{anafig2}. It improves over the radius approximation
but clearly Eq. (\ref{FF}) is softer than a dipole.

In conclusion,
we have shown results for the anapole form factor of the nucleon 
in leading order in $\chi$PT.
We have not yet examined extensively the form factor in next order.
There undetermined short-range isoscalar and isovector parameters appear,
and the anapole moment can no longer be predicted in a model-independent
way. 
However, the momentum dependence continues to be determined
by the pion cloud ---with nucleons
and deltas in internal lines--- 
and an improved low-energy theorem can be derived.
Under the assumption that higher-order results are afflicted by
the same dimensionless factors seen above, 
the error in Fig. \ref{anafig2}
at momentum $Q$ would be $\sim Q/m_\rho$.
Within such an error, the approximation
of using the anapole moment instead of the full form factor
would be good to about 15\% in the SAMPLE experiment,
but larger changes appear at momenta relevant to the JLab experiments.
On the other hand, if in next order
the dimensionless factors that affect the $Q^2$ variation
turn out closer to 1 than to 1/5, 
the momentum dependence from sub-leading order could be 
(accidentally) comparable to the leading order considered above.
Only an explicit calculation can verify if
the contributions that are formally subleading
are indeed numerically small.
We are currently investigating these issues \cite{nois}.

\vspace{1cm}
\noindent
{\large\bf Acknowledgements}

\noindent
We thank Bob McKeown and the group at the Kellogg Lab
for getting us interested in this problem 
and for comments on this work,
Paulo Bedaque for a useful suggestion,
and Wick Haxton and Barry Holstein
for pointing out the unusual convention for the sign of
$\gamma_5$ in DDH. 
CMM acknowledges a fellowship from FAPESP (Brazil),
grant 99/00080-5. 
This research was supported in part by the NSF grant PHY 94-20470.

\noindent
{\bf Note added}

\noindent
After this work was completed, it was pointed out to us that the %
nucleon form factor in leading order had previously been calculated in
\cite{deutana}.


\vspace{1cm}
                  
\begin{thebibliography}{50}
\bibitem{sample} 
D.T. Spayde et al (SAMPLE Collaboration), {\tt nucl-ex/9909010};
B. Mueller et al (SAMPLE Collaboration), 
{\it Phys. Rev. Lett.} {\bf 78} (1997) 3824.

\bibitem{happex} K.A. Aniol et al (HAPPEX Collaboration), 
{\it Phys. Rev. Lett.} {\bf 82} (1999) 1096.

\bibitem{g0} G0 Collaboration, www.npl.uiuc.edu/exp/G0/G0Main.html

\bibitem{zeldovich} 
Ya.B. Zel'dovich, 
{\it Sov. Phys. JETP} {\bf 6} (1958) 1184; 
{\it Sov. Phys. JETP} {\bf 12} (1961) 777.

\bibitem{holmus} M.J. Musolf and B.R. Holstein, 
{\it Phys. Rev.} {\bf D43} (1991) 2956;
W.C. Haxton, E.M. Henley, and M.J. Musolf,
{\it Phys. Rev. Lett.} {\bf 63} (1989) 949.

\bibitem{kapsav} D.B. Kaplan and M.J. Savage,
{\it Nucl. Phys.} {\bf A556} (1993) 653.
 
\bibitem{ulfreview} 
V. Bernard, N. Kaiser, U.-G Mei{\ss}ner, 
{\it Int. J. Mod. Phys.} {\bf E4} (1995) 193.

\bibitem{weinberg} 
S. Weinberg, {\it Phys. Lett.} {\bf B251} (1990) 288; 
             {\it Nucl. Phys.} {\bf B363} (1991) 3.

\bibitem{hpiNN} W. Haeberli and B.R. Holstein, in 
{\it Symmetries and Fundamental Interactions in Nuclei},
W.C. Haxton and E.M. Henley (eds.), World Scientific, Singapore (1995);
W.T.H. van Oers, {\tt hep-ph/9910328}.

\bibitem{DDH} B. Desplanques, J.F. Donoghue, and B.R. Holstein,
{\it Ann. Phys.} {\bf 124} (1980) 449.

\bibitem{nois} C.M. Maekawa, J.S. da Veiga, and U. van Kolck, in progress.

\bibitem{deutana} M. J. Savage and R. P. Springer, nucl-th/9907069.

\end{thebibliography}
\end{document}